\documentclass[fleqn,twoside]{article}
\usepackage{espcrc2}

\usepackage{graphicx}
\usepackage[figuresright]{rotating}
\usepackage{amssymb}

\newcommand{\AmS}{{\protect\the\textfont2
  A\kern-.1667em\lower.5ex\hbox{M}\kern-.125emS}}
\newcommand{\bq}{\begin{eqnarray}}
\newcommand{\eq}{\end{eqnarray}}
\newcommand{\eps}{\varepsilon}

\title{Infrared singularities in one-loop amplitudes}

\author{M. Assadsolimani\address[BERLIN]{Institut f{\"u}r Physik, Humboldt-Universit\"at zu Berlin, D - 12489 Berlin, Germany}, 
        S. Becker\address[MAINZ]{Institut f{\"u}r Physik, Universit{\"a}t Mainz, D - 55099 Mainz, Germany}, 
        Ch. Reuschle\addressmark[MAINZ] and 
        S. Weinzierl\addressmark[MAINZ]
}

\begin{document}

\begin{abstract}
In this talk we discuss a purely numerical approach to next-to-leading order calculations in QCD.
We present a simple formula, which provides a local infrared subtraction term for the integrand 
of a one-loop amplitude.
In addition we briefly comment on local ultraviolet subtraction terms and on the required deformation 
of the contour of integration.
\vspace{1pc}
\end{abstract}

\maketitle

\section{Introduction}

The experiments at the LHC are faced with a high QCD jet rate.
An accurate description of jet physics is therefore mandatory.
Although jet observables can rather easily be modelled at leading order (LO)
in perturbation theory, this description suffers several drawbacks.
A leading order calculation depends strongly on the renormalisation scale
and can therefore give only an order-of-magnitude estimate on absolute rates.
Secondly, at leading order a jet is modelled by a single parton. This is a very crude
approximation and oversimplifies inter- and intra-jet correlations.
The situation is improved by including higher-order corrections in perturbation theory.
Of particular phenomenological interest are next-to-leading order (NLO) corrections to 
multi-jet observables, where the number of jets is in the range of $3$ to $6$ or $7$.
This is not a simple task: The complexity of the calculation increases with the number of final state jets. 
There has been significant progress in the last years 
with these calculations, either based on traditional Feynman diagram techniques 
\cite{Dittmaier:2007wz,Dittmaier:2007th,Campbell:2007ev,Binoth:2009wk,Bredenstein:2009aj,Jager:2009xx,Lazopoulos:2007ix,Binoth:2008kt}
or on cut techniques 
\cite{Berger:2009zg,Bevilacqua:2009zn,Ellis:2009zw,Melnikov:2010iu}.

In this talk we discuss a third method which is based on a numerical integration for the loop integrals.
This by itself is not a new idea and has been discussed before \cite{Soper:1999xk,Kramer:2002cd,Nagy:2003qn,Nagy:2006xy,Gong:2008ww,Anastasiou:2007qb}.
What is new is a simple and compact formula which approximates the integrand of a one-loop QCD amplitude
in all infrared singular regions.
This approximation can be taken as a local counterterm and opens the door for an efficient numerical
implementation.
In addition we need local counterterms for the ultraviolet divergences and a deformation 
of the integration contour. We briefly comment on these points.

\section{The subtraction method}

The subtraction method \cite{Catani:1997vz,Phaf:2001gc,Catani:2002hc}
is widely used to render the real emission part of an NLO calculation suitable 
for a numerical Monte Carlo integration.
At NLO one has the following contributions:
\bq
\lefteqn{
\langle O \rangle^{NLO} 
 = } & & 
 \nonumber \\
 & &
 \int\limits_{n+1} O_{n+1} d\sigma^R + \int\limits_n O_n d\sigma^V 
 + \int\limits_n O_n d\sigma^C.
\eq
Here a rather condensed notation is used. $d\sigma^R$ denotes the real emission contribution,
whose matrix element is given by the square of the Born amplitudes with $(n+3)$ partons.
$d\sigma^V$ gives the virtual contribution, whose matrix element is given by the interference term
of the one-loop amplitude with $(n+2)$ partons with the corresponding
Born amplitude.
$d\sigma^C$ denotes a collinear subtraction term, which subtracts the initial-state collinear
singularities.
Taken separately, the individual contributions are divergent and only their sum is finite.
In order to render the individual contributions finite, such that the phase space integrations
can be performed by Monte Carlo methods, one adds and subtracts a suitable chosen piece:
\bq
\lefteqn{
\langle O \rangle^{NLO} =  
 \int\limits_{n+1} \left( O_{n+1} d\sigma^R - O_n d\sigma^A \right)
} & & \nonumber \\
 & &
 + \int\limits_n \left( O_n d\sigma^V + O_n d\sigma^C + O_n \int\limits_1 d\sigma^A \right).
\eq
The term $(O_{n+1}  d\sigma^R - O_n d\sigma^A)$ in the first bracket is by construction integrable over the
$(n+1)$-particle phase space and can be evaluated numerically.
The subtraction term can be integrated over the unresolved one-parton phase space.
Due to this integration all spin-correlations average out, but colour correlations still remain.
In a compact notation, the result of this integration is often written as
\bq
\lefteqn{
 d\sigma^C + \int\limits_1 d\sigma^A  
 = } & & \nonumber \\
 & &
 {\bf I} \otimes d\sigma^B + {\bf K} \otimes d\sigma^B + {\bf P} \otimes d\sigma^B.
\eq
The notation $\otimes$ indicates that colour correlations due to the colour charge operators
${\bf T}_i$ still remain.
The terms with the insertion operators ${\bf K}$ and ${\bf P}$ pose no problem for a numerical evaluation.
The term ${\bf I} \otimes d\sigma^B$ lives on the phase space of the $n$-parton configuration and has the appropriate
singularity structure to cancel the infrared divergences coming from the one-loop amplitude.
Therefore $d\sigma^V + {\bf I} \otimes d\sigma^B$ is infrared finite.

The virtual correction term $d\sigma^V$ contains the one-loop amplitude and is usually considered
to be the bottleneck of a NLO calculation.
$d\sigma^V$ is given by
\bq
 d\sigma^V & = & 2 \;\mbox{Re}\; \left(\left.{\cal A}^{(0)}\right.^\ast {\cal A}^{(1)} \right) O_n d\phi_n
\eq
${\cal A}^{(1)}$ denotes the renormalised one-loop amplitude. It is related to the bare amplitude by
\bq
\label{eq_one_loop}
 {\cal A}^{(1)} & = & {\cal A}^{(1)}_{bare} + {\cal A}^{(1)}_{CT}.
\eq
${\cal A}^{(1)}_{CT}$ denotes the ultraviolet counterterm from renormalisation.
The bare one-loop amplitude involves the loop integration
\bq
\label{integrand_one_loop}
{\cal A}^{(1)}_{bare} & = & \int \frac{d^Dk}{(2\pi)^D} {\cal G}^{(1)}_{bare},
\eq
where ${\cal G}^{(1)}_{bare}$ denotes the integrand of the bare one-loop amplitude.
Within our approach we extend the subtraction method to the integration over the virtual loop momentum $k$.
We rewrite eq.~(\ref{eq_one_loop}) as
\bq
\lefteqn{
 {\cal A}_{bare}^{(1)} + {\cal A}_{CT}^{(1)} 
 =  
 \left( {\cal A}_{bare}^{(1)} - {\cal A}_{soft}^{(1)} - {\cal A}_{coll}^{(1)} - {\cal A}_{UV}^{(1)} \right)
} & & \nonumber \\
 & &
 + \left( {\cal A}_{CT}^{(1)}  
 + {\cal A}_{soft}^{(1)} + {\cal A}_{coll}^{(1)} + {\cal A}_{UV}^{(1)} \right).
\eq
The subtraction terms ${\cal A}_{soft}^{(1)}$, ${\cal A}_{coll}^{(1)}$ and ${\cal A}_{UV}^{(1)}$
are chosen such that they match locally the singular behaviour of the integrand of ${\cal A}_{bare}^{(1)}$ in $D$ dimensions.
The expression in the first bracket can therefore be integrated numerically over the loop momentum $k$ in four dimensions.
The term ${\cal A}_{soft}^{(1)}$ approximates the soft singularities, ${\cal A}_{coll}^{(1)}$ approximates
the collinear singularities and the term ${\cal A}_{UV}^{(1)}$ approximates the ultraviolet singularities.
These subtraction terms have a local form similar to eq.~(\ref{integrand_one_loop}):
\bq
{\cal A}^{(1)}_{x} = \int \frac{d^Dk}{(2\pi)^D} {\cal G}^{(1)}_{x},
\;\;\;
 x = soft, coll, UV.
\eq
The building blocks of the subtraction terms are process-independent. When adding them back, we integrate analytically
over the loop momentum $k$. The result can be written as
\bq
\lefteqn{
 2 \;\mbox{Re}\; {\cal A}^{(0)}
   \left( {\cal A}_{CT}^{(1)} + {\cal A}_{soft}^{(1)} + {\cal A}_{coll}^{(1)} + {\cal A}_{UV}^{(1)}\right)^\ast 
                O_n d\phi_n
 } & & 
 \nonumber \\
 & = & 
 {\bf L} \otimes d\sigma^B.
\eq
The insertion operator ${\bf L}$ contains the explicit poles in the dimensional regularisation parameter related
to the infrared singularities of the one-loop amplitude.
These poles cancel when combined with the insertion operator ${\bf I}$:
\bq
 \left( {\bf I} + {\bf L} \right) \otimes d\sigma^B & = &
 \mbox{finite}.
\eq
The operator ${\bf L}$ contains, as does the operator ${\bf I}$, colour correlations due to soft gluons.

\section{Colour decomposition}

It is convenient to decompose a full one-loop QCD amplitude into primitive amplitudes:
\bq
 {\cal A}^{(1)} & = & \sum\limits_{j} C_j A^{(1)}_j.
\eq
The colour structures are denoted by $C_j$ while
the primitive amplitudes are denoted by $A^{(1)}_j$.
In the colour-flow basis \cite{'tHooft:1973jz,Maltoni:2002mq,Weinzierl:2005dd} the colour structures are linear combinations of monomials in Kronecker $\delta_{ij}$'s.
Primitive amplitudes are defined as a colour-stripped 
gauge-invariant set of Feynman diagrams with a fixed cyclic ordering of
the external partons and a definite routing 
of the external fermion lines through the diagram \cite{Bern:1994fz}.

It is simpler to work with primitive one-loop amplitudes instead of a full one-loop amplitude.
Our method exploits the fact that primitive one-loop amplitudes have a fixed cyclic ordering of the
external legs and that they are gauge-invariant.
The first point ensures that there are at maximum $n$ different loop propagators in the problem, where $n$ is the 
number of external legs, while the second property of gauge invariance is crucial for the proof of the
method.
We therefore consider in the following just a single primitive one-loop amplitude,
which we denote by $A^{(1)}$, while keeping in mind that the full one-loop amplitude is just
the sum of several primitive amplitudes multiplied by colour structures.

We introduce some notation related to the primitive amplitude $A^{(1)}$.
Since the cyclic ordering of the external partons is fixed, there are only $n$ different propagators occurring
in the loop integral.
We label the external momenta clockwise by $p_1$, $p_2$, ..., $p_n$ 
and define $q_i=p_1+p_2+...+p_i$, $k_i=k-q_i$.
We can write the bare primitive one-loop amplitude in Feynman gauge as
\bq
\label{starting_point}
 A^{(1)}_{bare} & = & \int \frac{d^Dk}{(2\pi)^D} 
 G^{(1)}_{bare},
\nonumber \\
 G^{(1)}_{bare} & = &
 P(k) \prod\limits_{i=1}^n \frac{1}{k_i^2 - m_i^2 + i \delta}.
\eq
$G^{(1)}_{bare}$ is the integrand of the bare one-loop amplitude.
$P(k)$ is a polynomial in the loop momentum $k$.
The $+i\delta$-prescription instructs us to deform -- if possible -- the integration contour into the complex plane
to avoid the poles at $k_i^2=m_i^2$.
If a deformation close to a pole is not possible, we say that the contour is pinched.
If we restrict ourselves to non-exceptional external momenta,
then the divergences of the one-loop amplitude related to a pinched contour are either due to soft
or collinear partons in the loop.
These divergences are regulated within dimensional regularisation by setting the number of space-time dimensions
equal to $D=4-2\eps$.
A primitive amplitude which has soft or collinear divergences must have at least one loop propagator which corresponds
to a gluon. An amplitude which just consists of a closed fermion loop does not have any infrared divergences.
We denote by $I_g$ the set of indices $i$, for which the 
propagator $i$ in the loop corresponds to a gluon.

\section{Local infrared subtraction terms}

We can now present a formula for the local subtraction terms related 
to the infrared divergences \cite{Assadsolimani:2009cz}.
For simplicity we present here the formula for massless QCD.
The extension to massive particles is also known.
The infrared subtraction term can be written in unintegrated form as a soft and a collinear part:
\bq
\label{subtraction_term}
\lefteqn{
 G_{IR}^{(1)} =  G_{soft}^{(1)} + G_{coll}^{(1)},
 } & & \\
\lefteqn{
 G_{soft}^{(1)} =
 - 4 \pi \alpha_s i
 \sum\limits_{i \in I_g}
 \frac{4 p_i \cdot p_{i+1}}{k_{i-1}^2 k_i^2 k_{i+1}^2}  A^{(0)}_i,
 } & &
 \nonumber  \\
\lefteqn{
 G_{coll}^{(1)} =  
 - 4 \pi \alpha_s i
 \sum\limits_{i\in I_g}
 (-2) 
 \left( 
 \frac{S_i g_{UV}\left(k_{i-1}^2,k_i^2,\mu_c^2\right) }{k_{i-1}^2 k_i^2}
 \right. 
} & &
 \nonumber \\
 & & 
 \left.
+
 \frac{S_{i+1} g_{UV}\left(k_{i}^2,k_{i+1}^2,\mu_c^2\right) }{k_{i}^2 k_{i+1}^2}
 \right)
 A^{(0)}_i.
 \nonumber
\eq
The Born partial amplitude $A^{(0)}_i$ depends on the external momenta, but not 
on the loop momentum.
The constants $S_i$ are given by $S_q=S_{\bar{q}}=1$ and $S_g=1/2$, the index $i$ of $S_i$ refers to the external particles.
The function $g_{UV}$ ensures a regular behaviour of  the collinear term in the ultraviolet region.
A possible choice is \cite{Nagy:2003qn}
\bq
 g_{UV}\left(k_{i-1}^2,k_i^2,\mu_c^2\right) =
 \frac{1}{2} \left( \frac{-\mu_c^2}{k_{i-1}^2-\mu_c^2} + \frac{-\mu_c^2}{k_{i}^2-\mu_c^2} \right).
 \nonumber
\eq
$\mu_c$ is an arbitrary scale.
Integrating the soft and the collinear part we obtain
\bq
\label{integrated_ir_subtraction}
\lefteqn{
 S_\eps^{-1} \mu^{2\eps} \int \frac{d^Dk}{(2\pi)^D} G_{soft}^{(1)} = 
 \frac{\alpha_s}{4\pi} 
 \frac{e^{\eps \gamma_E}}{\Gamma(1-\eps)}
} & & \\
 & & \times
 \sum\limits_{i \in I_g}
 \frac{2}{\eps^2} 
 \left( \frac{-2p_i\cdot p_{i+1}}{\mu^2} \right)^{-\eps}
 A^{(0)}_i
 + {\cal O}(\eps),
\nonumber \\
\lefteqn{
 S_\eps^{-1} \mu^{2\eps} \int \frac{d^Dk}{(2\pi)^D} G_{coll}^{(1)} = 
 \frac{\alpha_s}{4\pi} 
 \frac{e^{\eps \gamma_E}}{\Gamma(1-\eps)}
} & & \nonumber \\
 & & \times
 \sum\limits_{i \in I_g}
 \left( S_i + S_{i+1} \right)
 \left( \frac{\mu_c^2}{\mu^2} \right)^{-\eps} \left( \frac{2}{\eps} + 2 \right)
 A^{(0)}_i
 \nonumber \\
 & &
 + {\cal O}(\eps).
 \nonumber
\eq
$S_\eps = (4\pi)^\eps e^{-\eps\gamma_E}$ is the 
typical volume factor of dimensional regularisation, $\gamma_E$ is Euler's constant  and $\mu$ is the renormalisation scale.
The conditions under which a single diagram leads to an infrared divergence
are well known 
\cite{Kinoshita:1962ur,Dittmaier:2003bc,Gluza:2007uw}.
The proof of eq.~(\ref{subtraction_term}) given in \cite{Assadsolimani:2009cz} 
uses then the fixed cyclic ordering and gauge invariance
to establish the relation at the level of primitive amplitudes.

Equation~(\ref{subtraction_term}) approximates the integrand of a primitive one-loop QCD amplitude 
in all soft and collinear limits.
The approximation is given by simple scalar two- and three-point functions, multiplied by a Born
partial amplitude.
One easily observes that the integrated form in eq.~(\ref{integrated_ir_subtraction}) 
agrees in the pole terms with the known result for the pole terms of a primitive one-loop 
amplitude.

Equation~(\ref{subtraction_term}) is a significant improvement with respect to the subtraction terms given in
ref.~\cite{Nagy:2003qn}.
In ref.~\cite{Nagy:2003qn} the subtraction terms where given graph by graph.
In contrast to this we see that eq.~(\ref{subtraction_term}) is formulated at the level of amplitudes and not at the level
of graphs.
The improvement results from the fact that an amplitude can be calculated efficiently using recurrence relations.
The recursive method is significantly faster compared to an approach based on individual diagrams.

A few remarks are in order:
Contrary to the subtraction terms for the real emission, there are no spin correlations
in the subtraction terms for the integrand of the one-loop amplitude. In the real emission case spin
correlations occur in the collinear limit.
This can be understood as follows:
From the proof of eq.~(\ref{subtraction_term}) one sees that in the collinear limit always one of the collinear gluons
carries an unphysical longitudinal polarisation. Hence, there are no correlations between two transverse polarisations
of a gluon.

Furthermore and also contrary to the subtraction terms for the real emission, there is no dependence on the variant
of dimensional regularisation (conventional dimensional regularisation, 't Hooft-Veltman scheme,
four-dimensional scheme) in the integrated result.
At first sight this is puzzling: 
In the real emission case the variant of dimensional regularisation introduces a scheme-dependent finite term.
On the other hand unitarity requires that this scheme-dependent finite term cancels in the final result.
It is instructive to investigate how this cancellation occurs.
The solution comes from the LSZ-reduction formula:
The renormalised one-loop amplitude with $n_g$ gluons, $n_q$ quarks and $n_{\bar{q}}$ antiquarks
is related to the bare amplitude by
\bq
\label{LSZ}
\lefteqn{
 {\cal A}_{ren}(p_1,...,p_n,\alpha_s)
 = 
 \left(Z_2^{1/2} \right)^{n_q+n_{\bar q}} \left( Z_3^{1/2} \right)^{n_g}
} & & \nonumber \\
& &  
 \times {\cal A}_{bare}\left(p_1,...,p_n,Z_g^2 S_\eps^{-1}\mu^{2\eps} \alpha_s\right).
\eq
$Z_g$ is the renormalisation constant for the strong coupling, given by 
\bq
 Z_g & = & 1 + \frac{\alpha_s}{4\pi} \left( - \frac{\beta_0}{2} \right)  \frac{1}{\eps} + {\cal O}(\alpha_s^2).
\eq
$Z_2$ is the quark field renormalisation constant and $Z_3$ is the gluon field renormalisation constant.
The LSZ reduction formula instructs us to take as field renormalisation constants the residue
of the propagators at the pole.
In dimensional regularisation this residue is $1$ for massless particles and therefore the field renormalisation
constants are often omitted from eq.~(\ref{LSZ}).
However $Z_2=Z_3=1$ is due to a cancellation between ultraviolet and infrared divergences \cite{Harris:2002md}.
In Feynman gauge we have
\bq
 Z_{2} & = & 1 + \frac{\alpha_s}{4\pi} C_F \left( \frac{1}{\eps_{IR}} - \frac{1}{\eps_{UV}} \right) + {\cal O}(\alpha_s^2),
 \nonumber \\
 Z_{3} & = & 1 + \frac{\alpha_s}{4\pi} \left( 2 C_A -\beta_0 \right) \left( \frac{1}{\eps_{IR}} - \frac{1}{\eps_{UV}} \right) 
 \nonumber \\
 & &
 + {\cal O}(\alpha_s^2).
\eq
Here we indicated explicitly the origin of the $1/\eps$-poles.
These poles introduce scheme-dependent finite terms of ultraviolet and infrared origin with opposite sign.
Now the cancellation of the scheme-dependent finite term of infrared origin is as follows:
The scheme-dependent finite term from the real emission contribution cancels with the scheme-dependent finite term
of infrared origin from the renormalisation constants.
This leaves the scheme-dependent finite terms of ultraviolet origin in the renormalisation constants and in the bare
one-loop amplitude.
These remain and give the result of the calculation in the chosen scheme.
One can convert from one scheme to another by a finite renormalisation.
We remark that
the scheme-dependence of the bare one-loop amplitude is entirely of ultraviolet origin.

\section{Local ultraviolet subtraction terms}

A primitive one-loop amplitude contains, apart from infrared divergences, also ultraviolet divergences.
In an analytic calculation regulated by dimensional regularisation these divergences manifest 
themselves in a single pole in the dimensional regularisation parameter $\eps$.
If we would do the calculation within a cut-off regularisation we would find a logarithmic dependence on the cut-off.

Within the numerical approach proposed here we need subtraction terms, which approximate the ultraviolet behaviour of the
integrand locally.
We first note that the one-loop amplitude has in a fixed direction in loop momentum space up to quadratic
UV-divergences.
These are reduced to a logarithmic UV-divergence only after angular integration.
For a local subtraction term we have to match the quadratic, linear and logarithmic divergence.
The ultraviolet subtraction terms have the form of propagator and vertex counterterms.
(Diagrams with five or more loop propagators are never UV-divergent.)
We can choose them such that after integration they are proportional to their tree-level counterpart.
This ensures that the sum of all integrated UV subtraction terms is again proportional to a tree amplitude.
As an example we give here the ultraviolet subtraction term for the quark-photon vertex.
In unintegrated form this term reads \cite{Beckerandmore}
\bq
\lefteqn{
 V_{qq\gamma} =    
 i Q_q e g^2 S_\eps^{-1} \mu^{4-D} 
} & & \\
 & &
 \int \frac{d^Dk}{(2\pi)^Di} 
 \frac{2 \left(1-\eps \right) \bar{k}\!\!\!/ \gamma^\mu \bar{k}\!\!\!/ + 4 \mu_{UV}^2 \gamma^\mu}{\left(\bar{k}^2-\mu_{UV}^2\right)^3}.
 \nonumber 
\eq
$\mu_{UV}$ is an arbitrary scale. 
The term proportional to $4\mu_{UV}^2 \gamma^\mu$ in the numerator is not divergent, but ensures that the integrated
expression has a simple form.
Integration leads to 
\bq
 V_{qq\gamma} =  
 - i \frac{Q_q e g^2}{(4\pi)^2} \gamma^\mu 
 \left( \frac{1}{\eps} - \ln \frac{\mu_{UV}^2}{\mu^2} \right)
 + {\cal O}(\eps).
\eq
Similar expressions can be found for all other UV-divergent self-energy and vertex corrections.
A possible choice of UV-subtraction terms has been given in \cite{Nagy:2003qn}.

\section{Contour deformation}

Having a complete list of ultraviolet and infrared subtraction terms at hand, we can ensure that the integration
over the loop momentum gives a finite result and can therefore be performed in four dimensions.
However, this does not yet imply that we can simply or safely integrate each of the four components of the loop momentum $k^\mu$
from minus infinity to plus infinity along the real axis.
There is still the possibility that some of the loop propagators go on-shell for real values of the loop momentum.
If the contour is not pinched this is harmless, as we may escape into the complex plane in a direction indicated by
Feynman's $+i\delta$-prescription.
However, it implies that the integration should be done over a region of real dimension $4$ in the complex space
${\mathbb C}^4$.

Let us consider an integral corresponding to a primitive one-loop amplitude with $n$ propagators minus the appropriate
IR- and UV-subtraction terms:
\bq
 I & = & 
 \int \frac{d^4k'}{(2\pi)^4} P(k')
 \prod\limits_{i=1}^n \frac{1}{{k_i'}^2 - m_i^2 + i \delta}.
\eq
$P(k')$ is a polynomial or more general a well-behaved function in the loop momentum $k'$.
The integration is over a complex contour in order to avoid -- whenever possible --
the poles of the propagators.
We set
\bq
\label{deformation}
 {k'}^\mu & = & k^\mu + i \kappa^\mu(k),
\eq
where $k^\mu$ is real. This introduces a Jacobian
\bq
 J(k) & = &
 \left| \frac{\partial {k'}^\mu}{\partial k^\nu} \right|,
\eq
and the integral becomes
\bq
 I & = & 
 \int \frac{d^4k}{(2\pi)^4} J(k) P(k'(k))
 \prod\limits_{i=1}^n \frac{1}{\left(k_i'(k)\right)^2 - m_i^2}.
 \nonumber 
\eq
The efficiency of the Monte Carlo integration depends crucially on the function $\kappa(k)$, which defines the deformation.
Possible choices have been discussed in \cite{Nagy:2006xy,Gong:2008ww}.

\section{Conclusions}

In this talk we discussed a numerical approach to NLO calculations in QCD.
Our main new result is a simple and compact formula, which approximates the integrand of a primitive
one-loop QCD amplitude in all limits where the loop integration leads to a soft or collinear singularity.
It thus serves as a subtraction term for the infrared divergences of a one-loop amplitude.
When combined with the corresponding subtraction terms for the ultraviolet divergences we may perform the loop
integration of the subtracted integrand in four dimensions.
It should be mentioned that there is still the need to deform the integration contour into the complex plane, 
since we have to avoid singularities of the integrand in regions where the integration contour is not pinched.


\end{document}